\documentstyle[11pt,newpasp,twoside,epsf]{article}
\def \cgs{erg~$\rm{s}^{-1}\rm{cm}^{-2}\rm{deg}^{-2}\;$}

\begin{document}
\title{Searching for Baryons with Chandra}
 \author{Lara Arielle Phillips and Jeremiah P. Ostriker}
\affil{Princeton University, Princeton, NJ 08544}

\begin{abstract}
At low redshifts, measurements of the total baryon content in stars, 
atomic and molecular hydrogen, and cluster gas fall a factor 
of two to four below the baryon density derived from
observed light-element ratios and nucleosynthesis 
arguments. A possible hiding place for a significant fraction of
the missing baryons 
is in the warm/hot gas at temperatures 
T$=10^{5}-10^{7}K$. We present predictions of the
contribution to the soft X-ray background from warm/hot gas 
emission calculated using new hydrodynamical simulations
 and discuss the possibility of 
detecting the spectral signature of this gas using the 
Chandra X-ray Observatory. 
\end{abstract}

\section{Introduction}

Lyman $\alpha$ measurements 
of the total baryon density (Rauch et al. 1998) 
at high redshift ($z \approx 2-3$) are in remarkable agreement 
with those derived from observed light-element ratios and
nucleosynthesis arguments (Burles \& Tytler 1998), which 
yield a value of $\Omega_{b,D/H} = 0.039 \pm 0.002$, assuming 
$h \equiv H_{0}/100 = 0.7.$ However, in the local universe, 
the combined observed contributions of stars, atomic
and molecular hydrogen, and cluster gas fall a factor of two to four 
below this number (Fukugita, Hogan, \& Peebles 1998). 

Recent simulations suggest that
at low redshift  
a large fraction of the baryons in the universe could be in 
the form of warm/hot gas 
in filaments (Cen et al. 1995; Cen \& Ostriker 1999; 
Dav\'{e} et al. 2000, and
references therein). This diffuse gas is also known as the
warm/hot intergalactic medium (WHIM). It is 
located outside of clusters of galaxies and is heated to
temperatures, $T = 10^{5}-10^{7}$K, intermediate between 
those of the hot cluster gas and the warm gas in voids. 
The comparatively low density
and temperature of the WHIM make it a challenge to
detect and as yet, attempts to observe it emission have yielded marginal
or negative results
(Briel \& Henry 1995; Wang, Connolly \& Brunner 1997; Boughn 1998;
Kull \& B\"ohringer 1999; Scharf et al. 2000).

As part of on-going work to characterize the X-ray 
background from the WHIM, we have calculated 
the integrated $0.2-10.$ keV spectrum from
this gas (Phillips, Ostriker, \& Cen 2000). 
We compare predicted flux levels
with existing limits from observations.
We have also generated model spectra to estimate
 the possibility of detecting the signature of this
gas using the high spatial resolution, 
energy resolution, and soft X-ray detection capabilities 
of the ACIS-S chips on the Chandra X-ray Observatory (Phillips 2000).

\section{The Integrated X-ray Background Spectrum 
from the WHIM}

The integrated X-ray background spectrum from the large
box was calculated by Cen \& Ostriker (1999) by adding the 
contribution from Orion-like stars, AGN and Bremsstrahlung emission from 
gas in their cosmological 
hydrodynamical simulation of a large $L=100h^{-1}$Mpc box.
The chosen cosmology is the "concordance model" (Wang et al.
2000), a flat low-density ($\Omega_{0} = 0.37$) universe with a 
cosmological constant ($\Omega_{\Lambda} = 0.63$),
$\Omega_{b} = 0.035$ and $H_{0} = 70$. 
However, the contribution from emission lines to 
the integrated spectrum was not included.

In order to determine the integrated X-ray spectrum from the WHIM
alone, we assume the gas is in thermal and ionization
equilibrium. We then use the Raymond-Smith (1977) code 
for optically thin plasmas, as modified by 
Cen et al. (1995), to obtain the average Bremsstrahlung and 
emission line spectra from the WHIM in the large box at 
output redshifts $z \leq 3$. These average spectra are 
used to obtain the integrated WHIM X-ray background contribution.
The AGN spectrum is estimated by normalizing the integrated  
AGN spectral template obtained from a smaller box so that
 at high energies ($> 30$ keV) the X-ray background is 
almost completely resolved into AGN (Mushotzky et al. 2000).
A more detailed description of these calculations is given in Phillips,
Ostriker, \& Cen (2000).
Since the contribution from stars falls off 
rapidly at energies greater than 0.2 keV, we can recover the total 
Cen \& Ostriker (1999) integrated spectrum by adding the AGN
contribution to the WHIM Bremsstrahlung spectrum.
The sum of the contribution from the WHIM and the AGN spectrum 
is used as the total simulated XRB spectrum in the discussion below.

\begin{figure}
\plotone{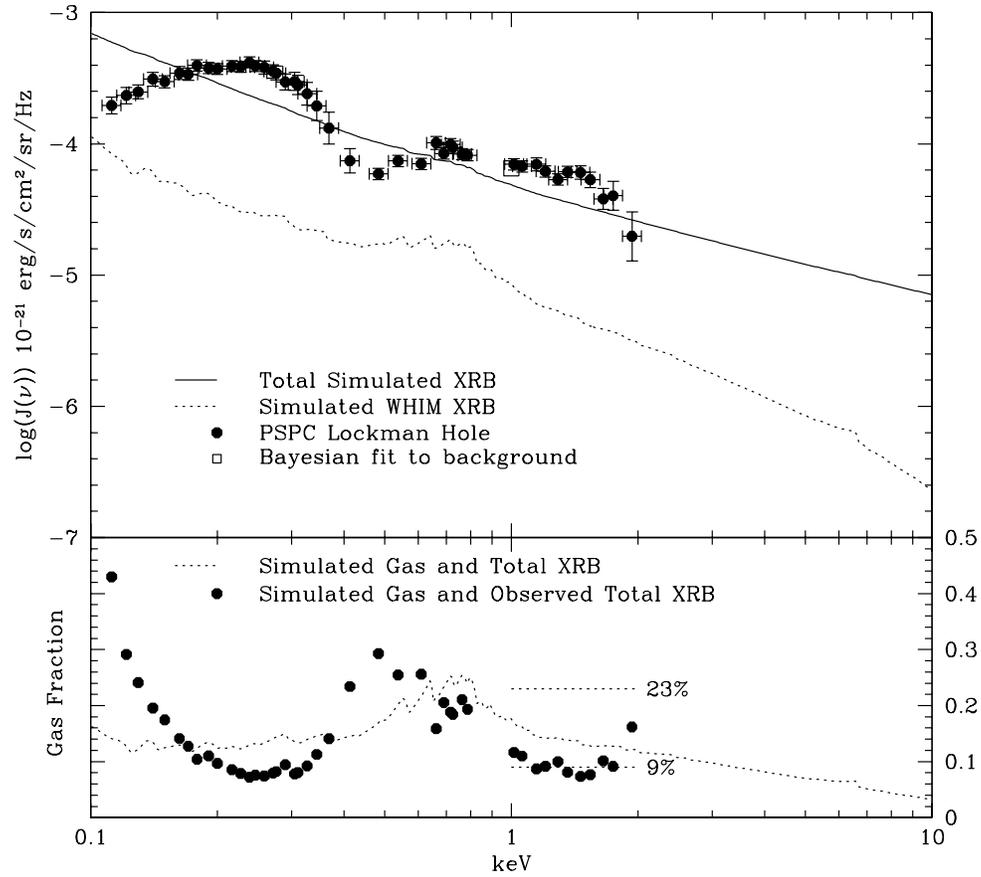}
\caption{Integrated XRB from AGN, stars and gas (solid line)
and from the WHIM alone (dotted line). The Miyaji et al. (1998)
Lockman Hole ROSAT PSPC XRB (circle) and Barcons et al. (2000) Bayesian fit
to XRB observations (square) are also shown. 
The fractional contribution of the WHIM
to the simulated total (line) and Miyaji et al. (1998) observed background
(symbols) is plotted in the lower panel.}
\end{figure}

The integrated spectrum from the WHIM is shown in
Figure 1 and is used in the following section to obtain
flux predictions to compare with observational limits.
The upper panel depicts our WHIM results 
along with the total X-ray background spectra from 
the large box simulation and from the Lockman Hole 
ROSAT PSPC data in Miyaji et al. (1998). The latter is consistent
with a Bayesian fit to the observed X-ray background at 1 keV 
by Barcons, Mateos, \& Ceballos (2000).
The predicted fraction due to the WHIM is shown 
in the lower panel as a function of the mid-range, 
Miyaji et al. (1998)
observed background (dots) and as a function
of the total integrated X-ray background from simulations (solid line).

\section{The X-ray Background Breakdown}

Table 1 shows the current percentage contributions from AGN and other 
sources to the X-ray background for extremum values of the X-ray
background (Hasinger et al. 1998, Mushotzky et al. 2000; Giacconi et al. 
2000). 
The low value for the X-ray background of 
$3.7 \times 10^{-12}$ \cgs
is taken from Gendreau et al. (1995),
the mid-range values from spectral fits to ROSAT PSPC data in 
Miyaji et al. (1998)
 and the upper limit of $ 4.4 \times 10^{-12}$ \cgs from 
Chen, Fabian \& Gendreau (1997). In all cases, the errors shown are estimates
obtained from adding errors stated in the above papers in quadrature.
 The residual fraction of the background is given in the
last row and varies from  9\% to 23\%. In the $1-2$ KeV band, 
at most $10^{-12}$ \cgs remains to be resolved (Mushotzky et al.
2000). 
It is unlikely that this will all be resolved into AGN however, 
since fainter AGN have harder spectra and the high spatial resolution 
(0.5 arcsec) of Chandra ensures that most of the sources have now been
resolved. Integrating the WHIM spectrum we obtained in the previous
section over the energy band $1-2$ keV, we obtain a value of 
$0.22  \times 10^{-12}$ \cgs which represents $9\%$ of the
simulated background and is
consistent with the limits placed on the contribution by observations.

\begin{table}
\caption{Sources in 1 to 2 keV X-ray Background}
\begin{tabular}{llllr}
\tableline
Reference & S \tablenotemark{1,2}
& Resolved \tablenotemark{1,3} 
& Remainder \tablenotemark{1}
& \% \tablenotemark{4} \\
\tableline
Gendreau et al. & $3.7 \pm 0.53$ & $3.38 \pm 0.014$ & $0.32 \pm 0.54$ & 9. \\
Miyaji et al. & $4.2 \pm 0.33$ & $3.38 \pm 0.014$ & $0.82 \pm 0.34$ & 20. \\
Chen et al. \tablenotemark{5} & $4.4$ & $3.38 \pm 0.014$ & $1.12$ & 23. \\
\tableline
\tableline
\tablenotetext{1}{in $10^{-12}$ \cgs}
\tablenotetext{2}{total surface brightness in 1 - 2 keV band}
\tablenotetext{3}{Hasinger et al. (1998) and Mushotzky et al. (2000)}
\tablenotetext{4}{of XRB not resolved}
\tablenotetext{5}{error information not available}
\end{tabular}
\end{table}

\section{Detecting WHIM with Chandra's ACIS-S} 

We have simulated spectra obtained from adding all the WHIM emission
in a $8'\times8'$ field of view. 
A distribution of typical resulting fluxes is
 shown in Figure 2. The dotted line corresponds to the flux limit for
extended sources in deep Chandra exposures above which
the count rate in the ACIS-S chips
should allow us to use the
intermediate energy resolution of the instrument to study the
broad band spectral signature of the WHIM (Phillips et al.
1999). We plan to use the high spatial resolution of Chandra
to look at the background between the sources detected
in the the deep ACIS-S observation of the Hawaii Deep Survey
field SSA31 field (Mushotzky et al. 2000).
We expect the residual background emission to have spectral 
properties similar to 
those of the WHIM (Phillips et al. 2000b). 
The spectra and their implications will be described in greater detail 
in Phillips (2000).

\begin{figure}
\plotone{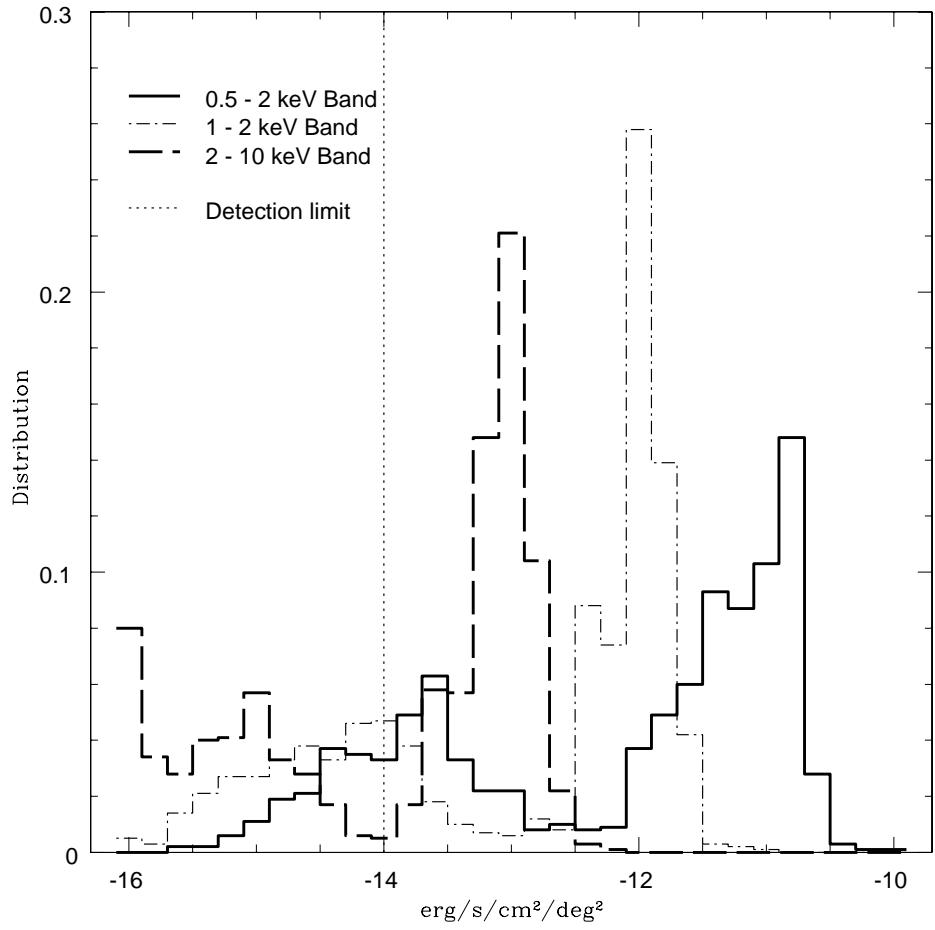}
\caption{Distribution of expected flux from the WHIM in $8' \times 8'$
FOV.}
\end{figure}

\section{Conclusion}

We have shown that the integrated X-ray background spectrum from
the WHIM we obtain from a large $100h^{-1}$ Mpc
hydrodynamical simulation is consistent with current
observational limits placed on the contribution from diffuse
gas to the X-ray background.
Our preliminary simulations of the spectrum one could obtain 
in a $8'\times 8'$ field of view indicate that it may be possible
to use ACIS-S chips on the Chandra X-ray Observatory to 
look between the sources and detect the spectral signature of
the WHIM. We are now working on a more detailed modeling of 
the integrated X-ray background from the WHIM 
in a $8' \times 8'$ field of view.
For a more detailed description and discussion
of the integrated background spectrum simulation outlined in 
this poster, please see Phillips, Ostriker, \& Cen (2000).

\acknowledgments

We wish to thank Renyue Cen his helpful comments.
This research was supported by NSF grants ASC-9740300 and AST-9803137.
LAP was supported in part by an award from the NSERC (Canada) 
and a grant from Zonta International.

\end{document}